# Patient-specific and physiological load sustaining synthetic graft substitutes for fusion of critically sized segmental bone defects

Rebecca Chung[1*], Dilhan M. Kalyon[1,2], and Antonio Valdevit[1]

Department of Biomedical Engineering[1]

Department of Chemical Engineering and Materials Science[2]

Stevens Institute of Technology

Castle Point on Hudson, Hoboken, NJ 07030, USA

*Corresponding Author (Current address):

Rebecca Chung

rebecca.chung@pennmedicine.upenn.edu

McKay Orthopaedic Research Laboratory

University of Pennsylvania

3450 Hamilton Walk, Philadelphia, PA 19104, USA

# Abstract

Critically sized defects are currently treated via autologous or allograft bone grafting, distraction osteogenesis, and membrane induction. However, these methods have major weaknesses which encourage the development of synthetic bone graft substitutes. A bioresorbable and physiologically load sustaining graft substitute was fabricated from polylactic acid using 3D printing, a versatile method which enables parameters to be customizable in a patient-specific manner. The internal architecture consists of vertical and horizontal conduits that are organized to achieve gradients in porosity and pore size in the radial direction to emulate cortical bone encapsulating cancellous bone. A diameter of 30mm and height of 10mm was designed to target the repair of critically sized long bone defects. The compressive properties of these graft substitutes were determined to be comparable to adult human femurs. Furthermore, dynamic fatigue testing at twice the human torso weight (800N) indicated that the bone graft substitute achieved stability as quickly as 200,000 cycles, suggesting its ability to support weight bearing loads while it gradually degrades as bone cells proliferate and mineralize. Additionally, the graft substitutes were seeded with human fetal osteoblasts to validate homogenous cellular attachment and cytocompatibility. Graft substitutes were also dynamically cultured for 28 days under various loading regimens using human mesenchymal stem cell differentiated osteoblasts, and these resulting tissue constructs were subjected to fatigue testing. The frequency of loading during dynamic culturing had a profound effect on the rate of mineralization which was particularly high at 2Hz, confirmed by Alizarin Red staining of the calcium deposits. Taken together, these bone graft substitutes have potential for use as orthopaedic implants to facilitate the repair of critically sized bone defects.

# 1. INTRODUCTION

Severe injuries to bone occur as a result of tumor resection, high energy traumas, or developmental deformities. Although bone tissue has remarkable self-healing capacities to remodel itself, spontaneous healing does not occur within a patient's life time for critically sized defects, for which the bone end gap is typically greater than 2-3 times the diaphysis diameter (Gugala et. al., 2007; Pneumaticos et. al., 2010). Current therapeutic approaches for the management of critically sized defects include autologous or allograft bone grafting, distraction osteogenesis, and membrane induction (Mauffrey et. al., 2015). As discussed next each of these currently available methods has major limitations, thereby encouraging the development of other alternatives including synthetic bone graft substitutes (Greenwald et. al., 2001; Hutmacher, 2000; Stevens, 2008; Roberts and Rosenbaum, 2012; Campana et. al., 2014).

Implementation of autologous bone is the gold standard for bone grafting and is recommended for voids that are less than 5 cm (Laurencin et. al., 2006). Autologous bone, generally removed from the iliac crest, is osteoinductive, osteogenic, nonimmunogenic and does not carry the risk of transmissible disease (Bauer and Muschler, 2000; Pneumaticos et. al., 2010). However, autologous bone is a finite resource, lacks a structural support, and its harvesting process involves morbidity associated with a second surgical site. Alternatively, cancellous allografts which are harvested from cadavers, can be used for repair of critically sized defects. In contrast to autologous grafts, allografts have unlimited availability, offer structural support, and their use eliminates donor site morbidity (Perry, 1999; Muscolo et. al., 2006). However, allografts lack osteogenic potential since mesenchymal stem cells, osteoblasts, or osteoclasts are unable to survive within the avascular environment. Their use can be further problematic in clinical practice because they present an increased risk of immunogenicity and disease transmission (Chmell et. al., 1995; Perry, 1999).

Distraction osteogenesis, also known as the Ilizarov method, uses the bone's natural capacity for regeneration under tension and involves attaching the segmental bone fragments to a circular external fixator with tensioned wires (Aronson, 1997). It has the disadvantage of relatively long treatment times, which can be as long as years, and carries the risks of infection, pin breakage and prolonged pain (Lasanianos et. al., 2009). The induced membrane method involves a two stage process (Masquelet and Begue, 2010), during which a bone cement, poly(methyl methacrylate) (PMMA), is implanted and stabilized at the defect site first. After 6-8weeks the bone cement becomes encapsulated into a fibrous membrane (Pelissier et. al., 2004). The fibrous membrane is incised to allow the removal of the PMMA spacer and an autologous or allograft implant is packed into the volume vacated by the spacer. The disadvantages of the induced membrane method include infection risk and the necessity of two surgeries (McConoughey et. al., 2015).

In comparison to these conventional methods, polymer based synthetic graft substitutes offer significant benefits in the management of critically sized bone defects (Coombes et. al., 1994; Khan, 2005; Kroeze et. al., 2009; Brydone et. al., 2010; Sheikh et al., 2015). With the appropriate selection of polymer, synthetic graft substitutes can be biocompatible, bioresorbable, and microporous with interconnected porosity (Chen et al., 2002; Pneumaticos et. al., 2010). Graft substitutes can be incorporated with various media, including hydrogels, which can carry mesenchymal stem cells and release bone growth factors in a time dependent fashion (Liu et. al., 2013). Such media can also carry endothelial cells and endothelial growth factors for vascularization (Shekaran et. al., 2014; Zigdon-Giladi et. al., 2015). Such synthetic graft substitutes can be rendered functionally graded with bioactive concentrations and porosity/mechanical property distributions to better mimic the complex hierarchical organization of native bone tissue (Erisken et. al., 2008; Ozkan et. al., 2010; Lu et. al., 2010; Erisken et. al., 2011; Ergun et. al., 2011a, Ergun et. al., 2011b; Ergun et. al., 2011c; Ergun et. al., 2012a; Ergun et. al., 2012b). Via such methods available for grading it is possible to alter the structure and properties of the bone graft substitute as a function of location in such a way to emulate the biofunctionality of native bone tissue, i.e., for example, the cortical bone encapsulating cancellous bone. Furthermore, polymer-based synthetic graft substitutes can be patient specific, i.e., the graft substitute can be tailored and fabricated to the geometry, bioactive and physical property needs of individual patients (Khan et. al., 2005). If the structure and the resulting mechanical properties of the bone graft substitute, especially for physiologically-relevant stresses, are tailored to biomimetically resemble those of native bone, then optimum load transfer would occur to promote integration (Guo, 2011; Stevens, 2008; Mercado-Pagan et. al., 2015).

The objective of our investigation was to develop a synthetic bioresorbable graft substitute in conjunction with an industrially-relevant fabrication method to be utilized as an orthopaedic surgical implant for the fusion of critically sized segmental bone defects. Fabricated from biocompatible poly-lactic acid, the graft substitute is comprised of a two-layer, i.e., a graded skin and core structure exhibiting varied porosity, pore size, and moduli in the radial direction to mimic the structural heterogeneity found in native bone tissue. The mechanical properties of the graft substitute were tailored to enable load transfer without failure under the high physiological loads that are typically experienced by adult human femur, and the restoration of biomechanical function is a major fundamental requirement in orthopaedic tissue engineering (Vunjack-Novakovic et. al., 2005; Liebscher and Wettergreen, 2003). The fabrication method selected, i.e., 3D printing, is sufficiently versatile for customized design and fabrication of the geometry and porosity distributions that are necessary for the treatment of critically sized defects of individual patients (Leong et. al., 2003; Leong et. al., 2008; Tarafder et. al., 2013; Mok et. al., 2016).

## 2. MATERIALS AND METHODS

### 2.1 Graft substitute design and fabrication

A multi-layered, skin/core bone graft substitute, which emulates cortical bone (the skin) encapsulating cancellous bone (the core), was fabricated from bioresorbable poly-lactic acid (PLA). The PLA was acquired in a filament form for 3D printing (Makerbot Industries, LLC, Brooklyn, NY) and converted into graft substitutes using a Makerbot Replicator 5th Generation 3D printer (Makerbot Industries, LLC, Brooklyn, NY). A skin/core graft substitute design that targets the repair of critically sized defects of long bones, is shown in Figure 1. The graft substitute measures 30mm in diameter for the skin layer, 20mm in the diameter for the core layer, and 10mm in height (Figure 1A). The design possesses an organized ringed lattice structure encompassing vertical and horizontal oriented conduits with gradually varied porosity from the core layer to the skin layer to emulate the cortical and cancellous bone distributions found in native bone tissue and to guide the tissue regenerative process (Figure 1B) (Di Luca, 2016). These vertical and horizontal conduits in the design mimic the plywood anatomy and structural heterogeneity of bone (Pompe et. al., 2003; Polo-Corrales, et. al., 2014; Reznikov et. al., 2014; Osterhoff, 2016; Le et. al, 2018), where the vertical channels contribute to sustain structural integrity under loading while the incorporation of horizontal channels enables ample fluid flow and an even distribution of external loads (Figure 1A-D). The average pore sizes of the core and skin layers are 1880±133 μm (Figure 1C) and 980±56 μm (Figure 1D), respectively and were selected to yield enhanced osteogenesis and sustained vascularization (VHX-5000 Digital Microscope, Keyence, Itasca, IL) (Karageorgiou and Kaplan, 2005; Di Luca, 2016). The targeted relatively higher modulus of the outer core layer should provide a stable mechanical protective casing to minimize the onset of potential stress shielding.

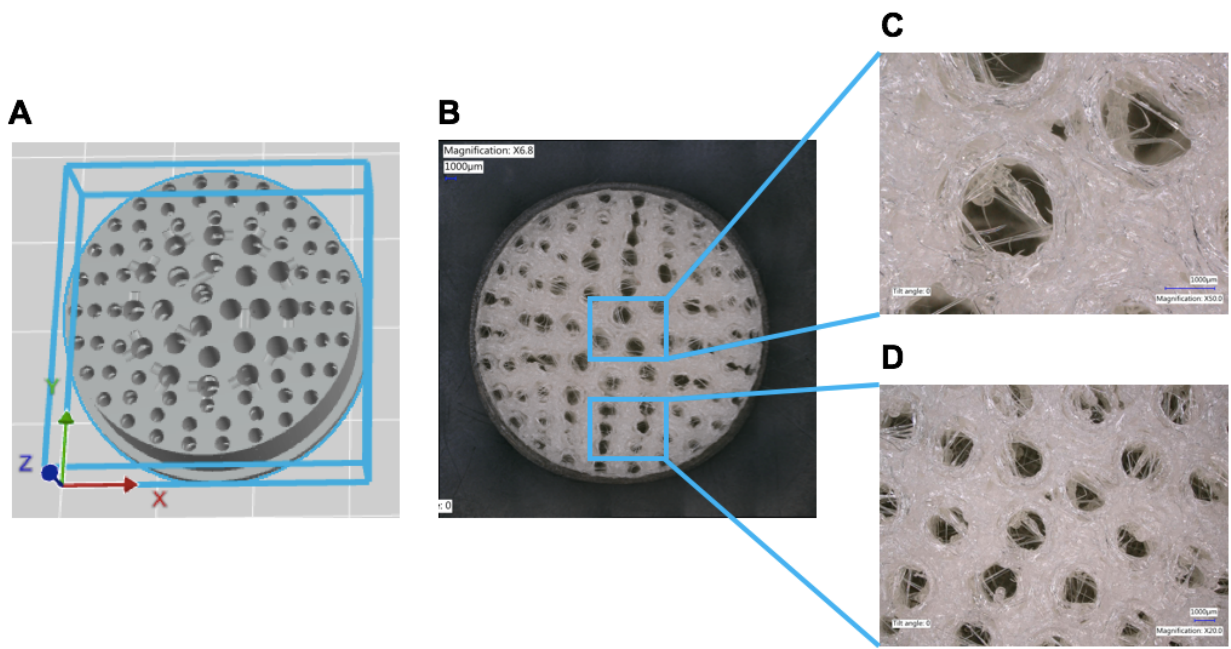

Figure 1. Graft substitute design utilized for investigations in this study. (A) AutoCAD rendering uploaded to 3D printer software in preparation for fabrication (B) Top-down view of whole graft substitute and enlarged (C) core layers and (D) skin layers imaged using Keyence VHX-500 Digital Microscope.

Modulus distributions in the radial direction are achieved by decreasing the number of horizontal and vertical conduits across from the inner core layer towards the outer layer of the graft substitute. The design is readily customizable using three-dimensional computer aided design software as demonstrated in Figure 1. The use of 3D printing technology allows the incorporation of multiple layers, i.e., for example the skin and core layers to mimic cortical bone encapsulating cancellous bone, as well as rendering the design to be patient specific by integrating the anatomical and geometric conditions of the defect sites of individual patients into the design (Bose et. al., 2013; Pobloth et. al., 2018). In clinical practice, a patient in need of bone graft surgery for segmental bone replacement traditionally undergoes dual-energy X-ray absorptiometry (DXA) and computed tomography (CT) scans to reveal detailed assessments of cortical and cancellous bone mineral density, geometry of the defect, and local variations of bone, all of which are useful for the pre-operative designing of a patient specific graft substitute using computer aided design software immediately amenable for 3D printing and nurtured in bioreactor systems until ready for implantable use (Kleerekoper and Nelson, 1997; Forhlich et. al., 2008; Johanson et. al., 2011; Fielding et. al., 2013; He et. al., 2015; Marro et. al., 2016).

**2.2 Biological Viability of graft substitutes**

The following procedure was used to test the biocompatibility of the bone graft substitute. First, the 3D printed graft substitutes were sterilized by immersing them in 70.0% isopropyl alcohol for two hours, followed by two hours of exposure to UV light. Then, they were washed three times with Dulbecco's Modified Phosphate Buffered Saline (ThermoFisher Scientific, Waltham, MA) and followed by three times with complete growth media, which was a 1:1 mixture of Ham's F12 Medium Dulbecco's Modified Eagle's Medium with 2.5 mM L-glutamine, supplemented with 10% fetal bovine serum and 1% streptomycin (ThermoFisher Scientific, Waltham, MA). In parallel, human fetal osteoblast cells (hFOB 1.19) (ATCC, Manassas, VA), an established cell line (Yen et al., 2007) were cultured in T-75 flasks and in a humidified incubator at 37°C buffered with 5% $CO_2$ and grown to confluency. Following trypsinization, the hFOB cells were seeded on the graft substitutes at a concentration of $8.0 \times 10^5$ cells/mL, totaling 6.5mL to ensure the entire graft substitute surface area was covered. The tissue constructs were maintained in culture for a week until they were fixed using 4% paraformaldehyde stained with 0.05% methylene blue. The cell attachment and penetration across the layers of the graft substitute were examined via microscopic imaging analysis using a VHX-5000 Digital Microscope (Keyence, Itasca, IL) to verify whether this design would potentially yield a biocompatible environment for cultivating bone growth.

In a second set of experiments human mesenchymal stem cell (hMSC) differentiated osteoblasts (PromoCell, Heidelberg, Germany) were seeded on the bone graft substitutes and dynamically cultured for a period or 28 days under various mechanical stimulation regimens after cell seeding to elucidate the effects of loading on osteoblastic mineralization; 0.5Hz, 2Hz, 5Hz, and control (N=6 per group). In preparation for cell seeding, the graft substitutes were sterilized using the same methods previously described. Confluent hMSCs were trypsinized and seeded onto each graft substitute at a concentration of $8.0 \times 10^5$ cells/mL, totaling 6.5mL. The hMSCs were induced to differentiate into osteoblasts using Osteogenic Differentiation Media, according to the manufacturer's protocol (PromoCell, Heidelberg, Germany). The tissue constructs were subjected to daily mechanical stimulation for 28 consecutive days. Following the completed loading regimens, the tissue constructs were subjected to fatigue testing for mechanical evaluation.

**2.3 Mechanical Testing: Static Analysis**

Mechanical evaluation of the graft substitutes was performed using ASTM Standard F2077-14 as reference. To characterize the static mechanical properties, the graft substitute (N=5) was aligned with the loading axis of an MTS 858 Bionix Testing Machine (MTS, Eden Prairie, MN) (Figure 2A) and compressively loaded at 25mm/min until failure was observed, as evidenced by a sudden drop in the load versus deformation curve (Figure 2B). Load versus displacement data were collected at a sampling rate of 40Hz. The load versus displacement curves for each graft substitute were analyzed to determine the yield load, failure load, and stiffness (Figure 2B). A one sample t-test comparison between mean experimental findings with (upper) mean reported literature values was performed to detect for any statistical differences. A confidence interval of 95% ($\alpha = 0.05$) was used for the analysis.

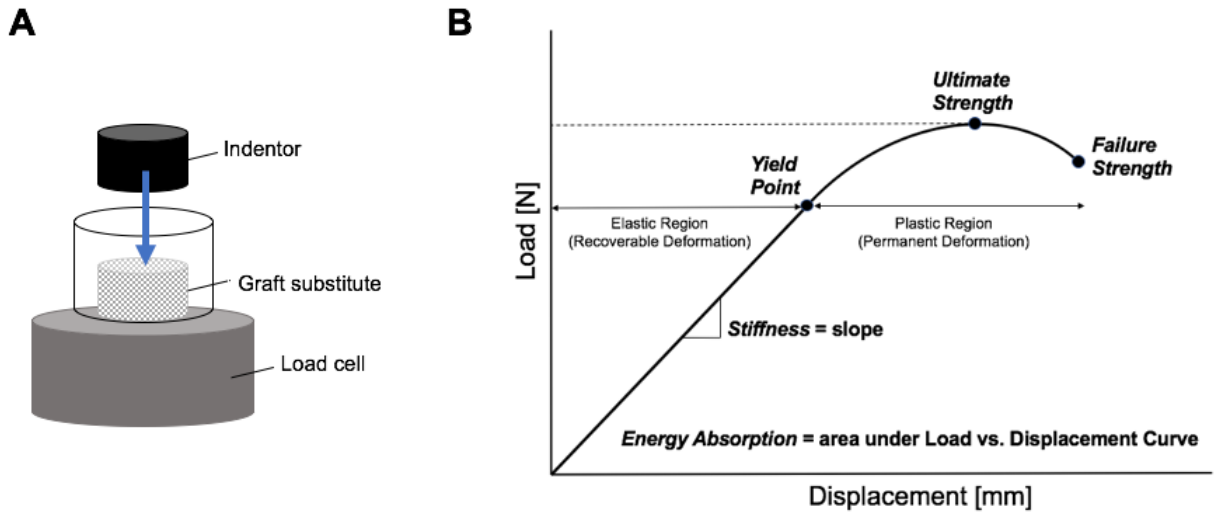

Figure 2. (A) Experimental test setup of bone graft substitute/bone specimen aligned with loading axis of MTS machine (B) Typical Load versus Displacement response of scaffold to static loading.

## 2.4 Dynamic Mechanical Testing: Fatigue Analyses

### 2.4a Data and statistical analyses of deformation versus cycle number curves

In fatigue testing, the nonlinear relationship between deformation versus cycle number curves produced may be expressed an exponential function in the form $Y=(Y0-Plateau)^{(-K*cycle\#)} + Plateau$ (Figure 3A). The resulting regression parameters produced consist of *Initial Deformation (Y0)*, *Rate (K)*, *Half-Life (ln(2)/K)*, or the number of cycles required to achieve a 50% change from *Y0,* Plateau (change in deformation over the course of the fatigue test) and *Span (Y0–Plateau)* may be statistically analyzed using a 1 way repeated measures ANOVA analysis followed by a Tukey post-hoc multiple comparison test to detect statistical differences (GraphPad Prism 5.0, San Diego, CA). A p-value less than 0.05 was used in the analyses for assessment of statistical significance. In the case of double (two-phase) decay fitting of nonlinear exponential regression as a result of the sum of a fast and slow exponential decay, the resulting regression parameters each produce a fast and a slow component (Figure 3B). The selection of a single or dual exponential function was based on an F-test with $p<0.05$.

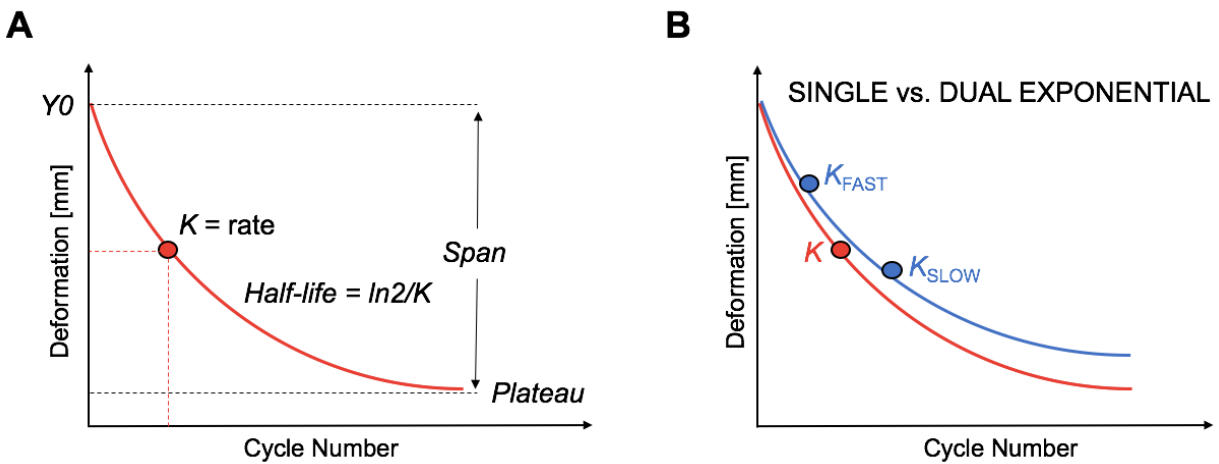

Figure 3. (A) The nonlinear relationship between deformation versus cycle number curve in fatigue mechanical testing may be expressed as an exponential function in the form Y=(*Y0-Plateau*)$^{(-K*cycle\#)}$+*Plateau* (B) Single (red) and dual (blue) exponential regression demonstrate how the presence of a two-phase exponential produces fast and slow components of curve parameters shown in (A).

**2.4b Dynamic Mechanical Testing: Bulk Fatigue Analysis**

The graft substitutes were bulk tested (N=6) under cyclic deformation using a Materials' Testing Machine (Bose ELF3200, Bose Corporation, Eden Prairie, MN). Samples were immersed in cell culture media 15min prior to loading to better simulate physiological conditions. Compressive sinusoidal fatigue loading was applied at low physiological loads, from -5N to -50N for 535 cycles. Continuous load vs. deformation data were acquired beginning at cycle number 10 and at subsequent 25 cycle intervals thereafter. Deformation, which occurs when there is a change from its original form under linear translation forces such as axial compression with respect to time may be elastic (recoverable change back to the original shape upon load removal) or plastic (non-recoverable). Following the fatigue test, the deformation change was determined by calculating the differences in displacement from -5N and -50N at each cycle count, which were then averaged for each graft substitute tested, then subjected to nonlinear exponential regression, followed by statistical analysis of the regression parameters (Prism 5.0, GraphPad Inc., San Diego, CA). Additionally, for comparison to native bone tissue behavior, porcine bone specimens with similar dimensions to the synthetic graft substitute were obtained by extracting the central core of thirty thoracic (T9-T11) vertebral bodies of 100kg porcine (Animal Technologies Inc., Tyler, TX) using a trephine and milled to achieve a flat surface for loading. These specimens were subjected to identical loading conditions and statistical analyses.

**2.4c Dynamic Mechanical Testing: Indentation Analysis**

To characterize the variations of mechanical properties as a function of location on the graft substitutes, indentation testing was performed. Prior to testing, the graft substitutes (N=6) were immersed in saline to simulate physiological conditions. Four sites on each sample surface (two at the outer skin layer and two at the inner core layer) were identified and subjected to 1005 cycles of loading at relatively low physiological loads varying between -10N to -100N, at a rate of 1Hz using a 4mm diameter indenter (Bose ELF 3200 Materials Testing Machine, Eden Prairie, MN). Continuous load vs. deformation data were acquired, beginning at cycle number 5 and at subsequent 50 cycle intervals thereafter. Deformation change at each cycle count was computed, averaged across each cycle count for each respective test site, averaged between respective inner and outer regions of the graft substitute, subjected to nonlinear analysis and to an F-test to determine whether a single or dual exponential would be the best fit (Prism 5.0, GraphPad Inc., San Diego, CA). Nonlinear regression parameters were statistically examined as previously described.

**2.4d Mechanical Testing: Fatigue Endurance Analysis**

The fatigue response of the graft substitutes (N=8) were examined using a Materials' Testing Machine (Bose ELF3300, Eden Prairie, MN). Pairs of graft substitute samples were subjected to compressive sinusoidal fatigue loading at -80N to 800N, -100N to -1000N, -140N to -1400N, and -180N to -1800N for a total of 5million cycles. Thus, the ratio of the minimum stress divided by the maximum stress resulted in an R ratio of 10. Continuous load vs. deformation data were acquired beginning at cycle number 500 and at subsequent 250,000 cycle intervals thereafter. The deformation change at each cycle count was computed, averaged across each cycle count for respective graft substitute loading pairs, and subjected to nonlinear exponential analysis as previously described (Prism 5.0, GraphPad Inc., San Diego, CA). Additionally, the change in percent strain and deformation over the course of the fatigue test was computed.

**2.4e Fatigue analysis of tissue constructs following 28-day mechanical stimulation**

The effects of mechanical loading of hMSC seeded tissue constructs under 28-days of daily mechanical stimulation (control, 0.5Hz, 2Hz, and 5Hz) on the rate of mineralization was also investigated. The fatigue analysis of the tissue construct obtained under mechanical stimulation conditions would provide insight on the progression of its mechanical properties as calcium is secreted to induce the mineralization of the construct. Thus, following the 28-day daily stimulation regimen, the tissue constructs (n=6 for each loading frequency group) were sacrificed and subjected to sinusoidal compressive loading at the adult body weight (or twice the adult torso weight) from -75N to -750N for 1005 cycles at a rate of 2Hz using a Materials' Testing Machine (Bose ELF3300, Eden Prairie, MN). Additionally, graft substitute samples which were not subjected to cell culture nor mechanical stimulation were also tested (n=6) using the parameters previously described to serve as controls. The deformation data produced from the mechanical testing was subjected a nonlinear regression of the data, followed by statistical analysis (GraphPad Prism 5.0, San Diego CA). A p-value less than 0.05 was used for assessment of statistical significance.

**3. RESULTS**

The bone graft substitute design comprised of a two-layer graded skin and core structure to mimic gradations found in native bone tissue (Leong et. al., 2008; Seidi et al., 2011; Di Luca et. al., 2016). The bone graft substitutes were fabricated via 3D printing, i.e., using a method that is industrially relevant and especially suitable for tailoring of the design to patient needs (Makerbot Replicator 5th Generation 3D Printer, Brooklyn, NY). The geometric size of the particular graft substitute design used for investigation reported here was based on the average transverse diameter of the midshaft of a tibia, as it is the most commonly fractured long bone in the human body due to lack of soft tissue coverage on the anteromedial surface which heightens the risk of bone loss (DeCoster et. al., 2004; Courtney et. al., 2011). The porosity of the graft substitute was graded in the radial direction, i.e. the graft substitute exhibited average pore sizes of 980±56 μm at the outer skin layer (Figure 1D) and 1880±133 μm at the inner core layer (Figure 1C). The average porosities at the core and skin layers of this graded structure ranged from 78% at the inner core layer to 55% at the outer skin layer. These porosity values were selected to reflect distributions found in native cancellous and cortical bone tissues (Rho et. al., 1998; Martin et. al., 2015). Previous investigations have shown pore size affects the progression of osteogenesis and graft substitutes intended for bone regeneration applications should possess pore sizes greater than 300 microns to yield enhanced new bone in-growth and formation of capillaries for sustained vascularization. (Freyman et. al., 2001; Karageorgiou and Kaplan, 2005). The interconnected porosities of the inner core and outer skin layers contributed by the vertical and horizontal conduits enable sustained biological transmission of fluid flow and solute transport between the skin and core layers which is essential for the maintenance of bone health (Mishra and Knothe-Tate, 2003).

The tissue constructs generated by the seeding and culturing of hFOBs in the graft substitutes were stained with methylene blue, which enabled the visualization of cellular attachment at the skin/core layers of the tissue constructs. Microscopic imaging analysis (Keyence VHX-5000 Digital Microscope, Itasca, IL) revealed that the osteoblasts migrated into both the skin and core layers and were attached throughout the volume of the entire graft substitute (Figure 4A-B). Thus, the bone graft substitute fosters a favorable environment to cultivate bone tissue proliferation, confirming its cytocompatibility.

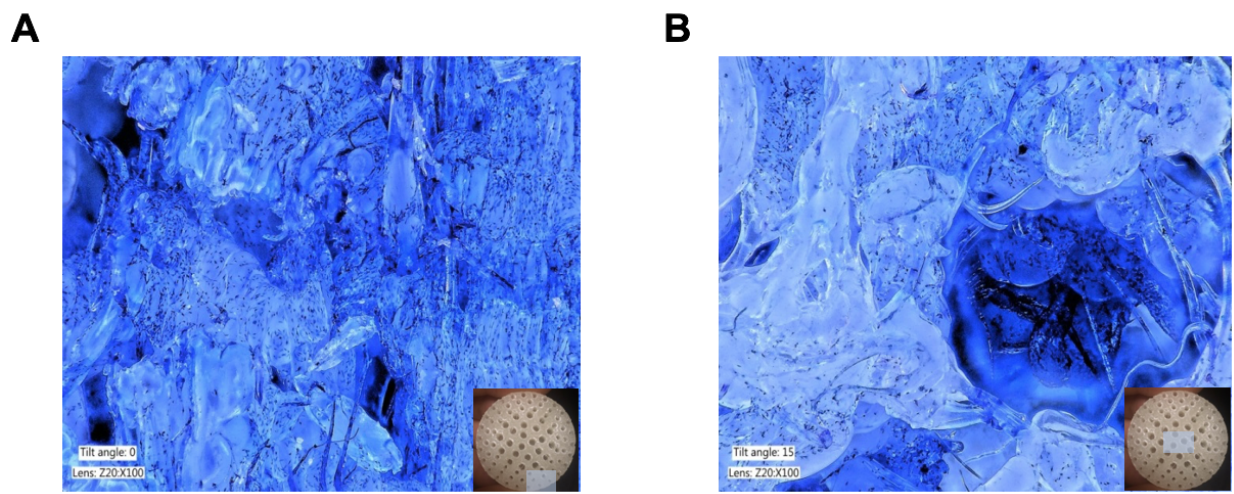

Figure 4. Methylene blue staining of tissue constructs at the (A) skin layers and (B) core layers using a VHX-5000 Digital Microscope (Keyence, Itasca, IL).

Statically, the average failure load and stiffness was calculated based on the load versus displacement curves generated from mechanical testing (Figure 2B). The compressive failure load of the graft substitute was determined to be (9645±54)N, which is comparable to the average reported failure load of the adult human femur, ranging from 7620N to 9076N (Smith et. al., 2014). Based on the failure load and surface area of the graft substitute, the ultimate stress was calculated to be 34.2 MPa. Furthermore, the stiffness of the graft substitute was determined to be (4024±304)N/mm, which is also comparable to the average stiffness associated with the adult human femur, ranging from 2924 N/mm to 4033 N/mm (Smith et. al., 2014). A one sample t-test comparison between experimental findings with (upper) mean literature values revealed no significant differences regarding stiffness ($p>0.05$) and a statistically elevated failure load for the graft substitutes ($p<0.001$).

Under bulk fatigue testing, the deformation versus cycle number curves for both the cancellous bone specimens (blue) and 3D printed graft substitutes (green) produced a non-linear relationship, which can be described using a single (one-phase) decay exponential function (Figure 5A). The observed relatively slow changes of the deformation values with increasing number of cycles for bone and graft substitute are also manifested in the relatively low and comparable *K*-values for bone and graft substitute. It should also be noted that the *K*-value of the graft substitute is similar to that of cancellous bone subjected to axial compression (Figure 5B) as no statistical significance was detected ($p>0.05$). The ability of the pores of the graft substitute to remain open during loading points demonstrate the potential of the graft substitute to encourage vascularization, which is a fundamental requirement for long term healing of the bone defect (Karageorgiou and Kaplan, 2005; O'Brien, 2011).

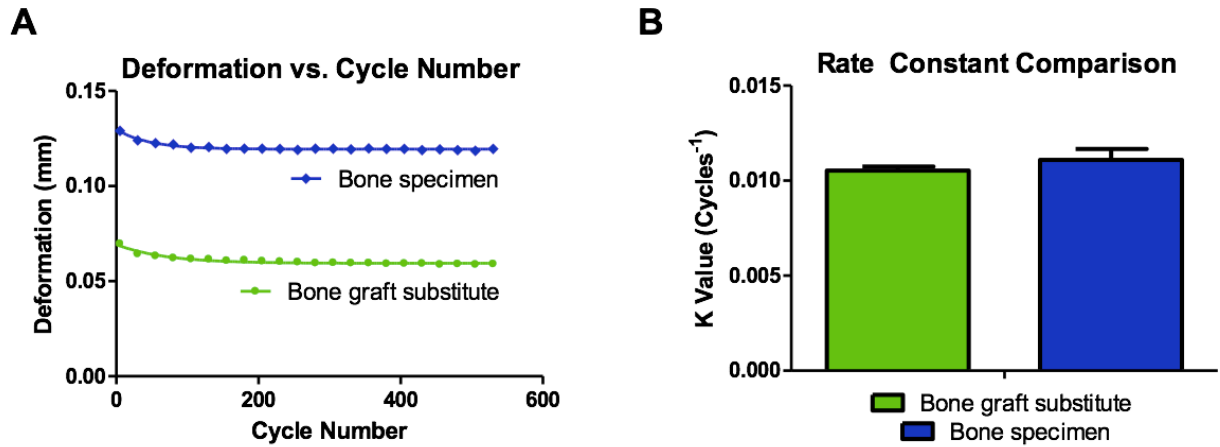

Figure 5. A) Resulting deformation versus cycle number B) *K*-value, or rate constant obtained from result of non-linear exponential regression of the graft substitute and trabecular bone in media.

Dynamic indentation testing enabled the mechanical properties of the skin and core layers of the bone graft substitutes to be characterized individually (Figure 6). The deformation fatigue curve analyses revealed a double (two-phase) exponential behavior for both the skin and core layers of the graft substitute (Figure 6A). The presence of a two-phase decay exponential generally suggests the biological behavior of the graft substitute where the outer skin layer contributes to a maintaining structural integrity as evidenced by a higher *PercentFast* in the outer skin layer (Figure 6B) *PercentFast* is the fraction of the *Span* (from *Y0 to Plateau*) accounted for by the faster of the two components, therefore a higher *PercentFast* attributes toward greater mechanical competence to maintain structure and strength which mimics the role of cortical bone (Ferretti, 1998; Rho et. al., 1998; Le et. al., 2018; Martin et. al., 2015).

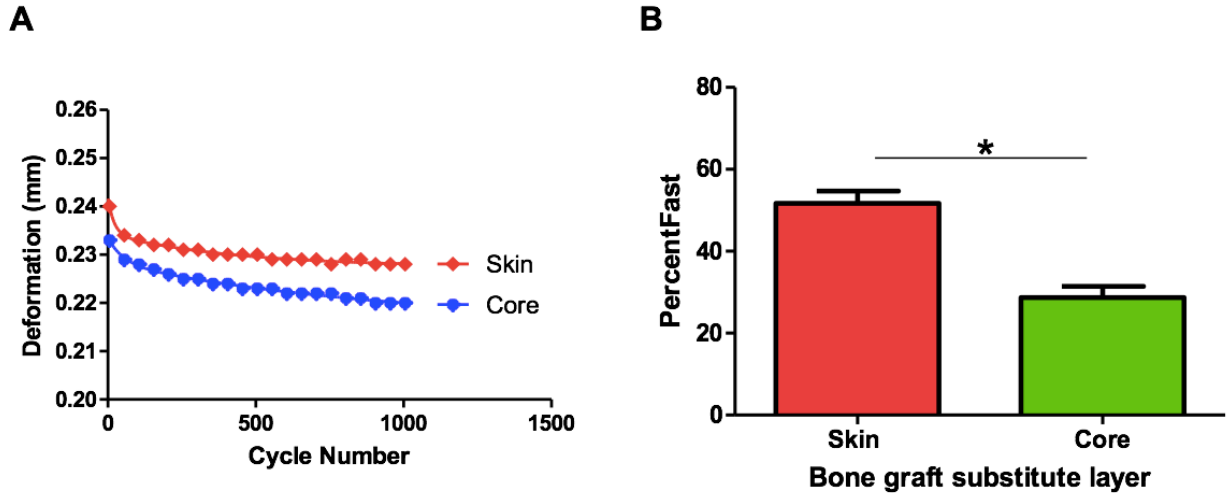

Figure 6. A) Resulting Deformation vs Cycle Number B) *PercentFast*, the fraction of the *Span* (from *Y0 to Plateau*) accounted for by the faster of the two components. *p<0.05

The fatigue endurance test further evaluated the suitability of the graft substitute as an implant for bone fusion (Figure 7). The fatigue behavior could be characterized via the use of single exponential decay of deformation versus cycle number at all load levels tested (Figure 7A). The strain at the final deformation at each load level was computed and the percent strain versus load was expressed as a nonlinear relationship (Figure 7B). All bone graft substitutes tested at the four load levels exhibited strain levels that were below 2%.

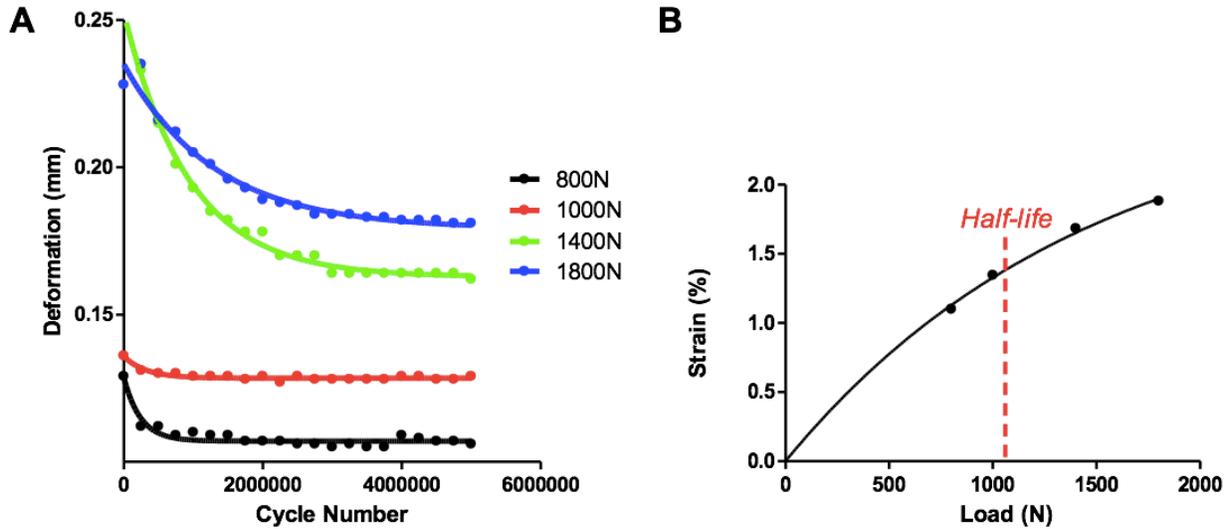

Figure 7. (A) Change in deformation over the course of the fatigue test (B) Non-linear exponential fit of the Percent Strain vs. Load curve

Bone graft substitutes must maintain sufficient strength from the moment of implantation into the patient until remodeling and fusion is complete. The results of the fatigue analysis of the bone graft substitutes are shown in Figure 7 where the data suggests that at approximately the adult body weight (or twice the torso weight) of 800N, the graft substitute achieves stability at approximately 200,000 cycles as evidenced by the rapid approach to *Plateau* behavior, which corresponds to a time point of approximately ten weeks post-operative surgery (Figure 8A). The *K*-values exhibited by the graft substitutes during the axial compression testing are similar to those of the cancellous bone specimens ($p>0.05$) (Figure 6A-B). As indicated earlier, matching of the important mechanical properties of the graft substitutes with those of the native tissues generates a more favorable healing process, especially for the long term (Stevens, 2008; Pobloth et. al., 2018).

Human mesenchymal stem cell differentiated osteoblasts seeded onto the graft substitutes to form tissue constructs underwent a 28-day duration of mechanical stimulation at various loading frequency regimens (control, 0.5Hz, 2Hz, 5Hz). Alizarin Red staining revealed a higher rate of mineralization at 2Hz, in comparison to 0.5Hz and 5Hz. Following the completion of this dynamic culture process, the tissue constructs were harvested and subjected to fatigue analysis. Sinusoidal compressive loading at the adult body weight (or twice the adult torso weight) was applied from -75N to -750N for 1005 cycles at a rate of 2Hz to characterize the progression of their mechanical properties over time as cells proliferate and mineralize toward healing. All of the nonlinear regressions of the experimental groups produced a dual (two-phase) exponential decay fitting of the deformation data (Figure 8A-D). The presence of a dual (two-phase) decay exponential can be interpreted as the manifestation of the elastic and viscous component responses of the viscoelasticity of the graft substitutes. The results of the nonlinear regression analysis suggest that there are both fast and slow components of the time dependent decay which reflect the physiological behavior of the graft substitute (Figure 3B).

Regarding *Y0*, the initial deformation, the comparison of all groups versus native graft substitutes were significantly different ($p<0.05$) (Figure 8A). A similar trend was observed with respect to plateau values (Figure 8B). With respect to *Half-life(fast)* values, with the exception of the control group all other groups were significantly different ($p<0.05$). The 2Hz loading frequency group exhibited significantly higher Half-Life values compared to those of the other groups. The relatively high *Half-life* value of the group of tissue constructs stimulated at 2Hz may be related to its high rate of mineralization response observed over the course of the 28-day dynamic culture (Figure 8C). Furthermore, the greater extent of mineralization of the tissue constructs stimulated at 2Hz also give rise to the highest *PercentFast* compared to all other groups as well as leading to increased stiffness observed and a lower number of cycles to reach plateau behavior (Figure 8D).

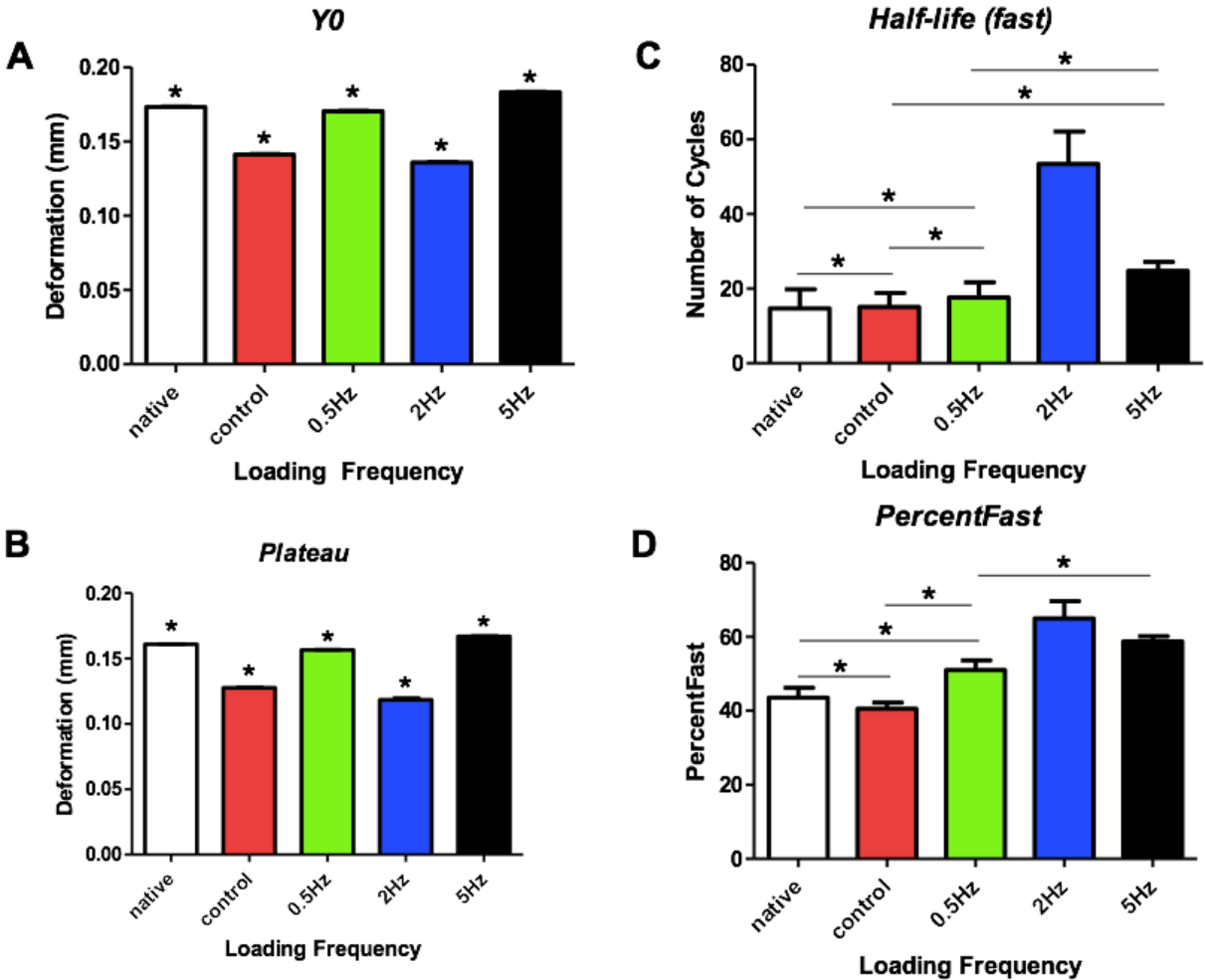

Figure 8. (A-D) Results of a 1-way ANOVA repeated measures test with Tukey post hoc test to statistically compare differences in loading frequency groups post 28-days of daily mechanical stimulation. *p<0.05

**4. DISCUSSION**

The need for synthetic graft substitutes that facilitate both mechanical function and tissue regeneration still remains a key hurdle in the field of regenerative medicine. There are still no treatment modalities that fulfill the fundamental requirements of matching mechanical and biological properties with native tissues. The development of tissue engineered human tissues pose challenges due to the complex structural and compositional gradations found in native tissues, and the interfaces of tissue-tissue transitions are difficult to mimic due to the limitations associated with conventional methodologies used for fabrication. Particularly with regards to bone, the dissimilarities of cortical and cancellous bone types require graft substitute designs which reflect these differences in bone morphology on a local level (Di Luca et. al., 2016). For example, in order to match the mechanical properties, graft substitutes for the repair of critically sized segmental bone defects must account for the structural heterogeneity found in bone by altering the porosity, pore size, and moduli along the transverse and axial directions (Osterhoff et. al., 2016).

Devices for bone fusion that are available on the market today specifically target spinal applications while synthetic bone graft substitutes targeting the repair and regeneration of critically sized defects in long bones are still needed. This present study represents one of the first developments to address the fusion of long bones on a patient-specific, load sustaining, and functionally graded basis (Figure 1). Previously, Bullens *et. al.* explored the concept of variable modulus by employing a Harm's cage, typically used in intervertebral body fusions, and filled it with morsellized bone graft chips to reconstruct segmental diaphyseal bone defects in goats. Examination of stability at the graft site revealed that the graft could be loaded by removing the load-transfer to the cage which facilitated the load transfer directly to the graft itself, thereby yielding the formation of healthy bone (Bullens et. al., 2009). Furthermore, the use of this treatment modality involving a cylindrical titanium mesh cage coupled with bone graft has also been reported by others (Cobos et. al., 2000; Ostermann et. al., 2002; Lindsey and Gugala, 2004;

Lindsey et. al., 2006). More recently, Pobloth *et. al.* developed titanium-mesh scaffolds with a honeycomb design to facilitate bone defect healing in sheep and found that lower stress shielding led to earlier defect bridging and advanced bony regeneration when coupled with a locking compression plate (Pobloth et. al., 2018). Such results emphasize the need for the development of functionally graded, physiologically load sustaining, and mechanobiologically functional bone graft substitutes for the fusion of long bones.

While advancements have been made in the development of functionally graded graft substitutes, the constraints with traditional fabrication methodologies do not satisfy current unmet needs. For example, such fabrication efforts have involved electrospinning of separate meshes and adjoining them under hydraulic pressure, layer-by-layer casting, freeze-drying, phase separation, and rapid prototyping techniques, including fused deposition modelling, selective laser sintering, and stereolithography. It has also been shown that graded polymeric implants and graft substitutes can be developed in an industrially-relevant manner utilizing (1) co-extrusion via a co-extrusion die, (2) hybrid extrusion and spiral winding method, and (3) hybrid twin screw extrusion and electrospinning methods. (Ozkan et. al., 2009; Ozkan et. al., 2010; Ergun et. al., 2011a; Ergun et. al., 2011b; Ergun et. al., 2011c; Ergun et. al., 2012a; Ergun et. al., 2012b; Erisken et. al., 2008a; Erisken et. al., 2008b; Erisken et. al., 2010; Erisken et. al., 2011). However, while these methods enable mass production, they are not particularly suitable for the fabrication of functionally-graded bone graft substitutes where customizations tailored to the specific needs of individual patients are necessary.

The utilization of patient-specific functionally-graded graft substitute designs can potentially alleviate physicians' challenges and concerns associated with tuning the ratios of cortical and cancellous bone in order to achieve the mechanical properties required for clinical application of the graft substitute (Kleerekoper and Nelson, 1997; Khan et. al., 2005;). Furthermore, as demonstrated here the use of additive manufacturing methods, such as the 3D printing method, enables rapid production of functionally-graded graft substitutes regardless of the complexity of the design (Kleerekoper and Nelson, 1997; Marro et. al., 2016; Pobloth et. al., 2018). It should be noted that the skin/core multilayered graft substitutes of this investigation could be fabricated within a 2-hour print time, at negligible operating and fixed costs.

The 3D printed graft substitutes were subjected to a series of mechanical tests for assessment and comparison with the mechanical properties of native bone tissue (Figure 2-3). The results of the static and fatigue analyses to mechanically evaluate this novel graft substitute design with its demonstrated compatibility (Figure 4) especially show promise for the fusion of critically sized segmental bone defects. Exhibiting an average failure load of (9645±54)N and average stiffness value of (4025±304)N/mm (Figure 2), both which are comparable to the adult human femur, this graft substitute can potentially provide the structural stability required for establishing a template to guide new tissue proliferation and growth while maintaining its mechanical properties to support the tissue regenerative process (Porter et. al., 2000; Giannoudis 2005; Ikada 2005; Wang et. al., 2008; Yu et. al., 2008). The low *K*-value parameters observed under bulk fatigue testing and greater *PercentFast* observed in the outer core layers of the graft substitute (Figure 5-6) under indentation fatigue testing signify the capability of tuning the mechanical properties of the graft substitute design to mechanobiologically optimize it to mimic the structure and functionality of native bone tissue behavior as well as tailor it to the specific needs of patients. The relatively low *K*-values suggest that the pores would provide ample fluid flow in and out of the graft substitutes which is favorable for cell migration and viability. The ability of the graft substitute pores to remain open during loading points demonstrate the potential of the graft substitute of encourage sustained vascularization, which is a fundamental requirement for long term healing of bone defects. (O'Brien, 2011).

Furthermore, the slower and gradual deformation changes versus the number of cycles applied based on the low value K parameters may be indicative of gradual mechanical changes due to increased fluid flow while the more rapid K values can be representative of the elastic phase within the graft substitute (Figure 6A). Such behavior is favorable for cell seeding as it leads to increased infiltration of cells, permeation of fluid flow, and nutrient exchange toward the inner core layers of the graft substitute under cyclic loading (Burg et. al., 2000). This confirms that mechanical properties of graft substitutes can be specifically designed to permit biological transmission between layers while mimicking native bone tissue properties.

Results of the fatigue endurance analysis revealed that the graft substitute achieved stability at approximately 200,000 cycles characterized by the rapid settling of *K*-values when tested at twice the torso weight of 800N which correlates to ten weeks post-surgery from a clinical perspective. As this is a comparable time frame at which patients achieve partial load bearing, this is an interesting finding as it suggests the minimization for use of conventional metal fixation systems such as locked compression plates which are traditionally employed to enhance and reinforce structural stability. As a result, healing in the absence of metal fixation systems alleviates the risk of postoperative complications such as bone resorption caused by stress shielding (Daniels et al., 1990; Uhthoff et. al., 2006; Ruggieri et. al., 2011) and reduces the need for an additional surgery to subsequently remove them. This minimizes the risk of postoperative complications which have been previously reported extensively in clinical practice (Sen and Miclau, 2007; Scheufler et. al., 2008; Graham et. al., 2010; Myeroff and Archdeacon, 2011; Flierl et al., 2013;). Furthermore, the bioresorbable PLA graft substitutes that are demonstrated in this investigation should provide the initial structural stability required to facilitate infiltration of tissue in-growth, while gradually degrading as bone tissue mineralizes toward healing (Porter et. al., 2000; Wang et. al., 2008; Yu et. al., 2008). For other applications, the degradation profile and new bone formation rates can be tuned to match for successful long-term incorporation (Ikada 2006; Polo-Corrales et. al., 2014). The type of polymer, its composition (i.e. formulation with bioactives), and structure can be tailored specifically to accommodate the intended application. The biodegradability aspect is essential as once the bone graft substitute is implanted, the surrounding tissues will absorb and metabolize it without the necessity to surgically remove the bone graft substitute when healing is complete.

The *Half-life*, based on the relationship: *Half-life=ln(2)/K*, was determined to be approximately (1017±11)N when the graft substitutes were subjected to fatigue endurance testing (Figure 7). This suggests that at a load that is 1.3 times greater than the adult body weight, the graft substitute is capable of maintaining the required mechanical strength to support the body and load bearing functions. This may be attributed to the horizontal channels in the bone graft substitute design which not only enable fluid flow throughout but also serve to distribute compressive load evenly. The strain$_{Half-life}$ was determined to be 1.4%, which is ideal since bone can withstand maximum strain values of approximately 2% (Figure 7B) (Osterhoff et. al., 2016).

It has been reported that use of mechanical loading in the development of tissue constructs may increase nutrient flow and yield a more even distribution of cell attachment (Duty et al., 2007). Such stimulation would improve bone mineralization to withstand mechanical stresses in accordance to Wolff's Law (Duty et al., 2007). Thus, the application of cyclical fatigue loading to the tissue constructs provided insight regarding the clinically-relevant response characteristics of tissue constructs obtained after cell proliferation and resulting mineralization achieved over 28-days of mechanical stimulation. The nonlinear analysis of the compressive indentation fatigue deformation data revealed the presence of a two-phase exponential decay behavior (Figure 3), which may be interpreted as having been developed as a result of elastic and viscous components in the viscoelastic response of the tissue construct. This two-phase decay behavior exhibits similarities to the viscoelastic behavior of native bone (Martin et. al., 2015; Le et. al., 2018). The elastic component is related to maintaining mechanical stability and structural integrity while the viscous component can be linked to the pertinent porosity that permits fluid penetration as well as nutrient and gas exchange.

The statistical analysis of the post-mechanical stimulation fatigue testing data revealed that the 2Hz loading frequency group possessed relatively higher *Half-life(fast)* values as compared to the other groups, which may be correlated to its high rate of bone proliferation and mineralization response over the course of the 28-day dynamic culture (Figure 8). This is indicative that in the dual (two-phase) exponential of the nonlinear regression, the "fast" or elastic component plays a greater role in maintaining structural support for the tissue constructs stimulated at 2Hz as compared to the other scaffold groups. This is valuable insight for the clinical implantation of these synthetic bone grafts in patients because 2Hz is the average stride frequency for humans, which suggests that walking while heading would reduce the risks of failure. Furthermore, the 2Hz loading frequency group possessed a relatively higher *PercentFast* as compared to the other groups. This indicates a relatively greater degree of mineralization, which perhaps reflects an increased stiffness thereby requiring a lower rate of net deformation change per cycle to achieve and maintain stability (Figure 8D). Studies have correlated the effects of implant stiffness on the osteogenic differentiation of mesenchymal stem cells, noting that cells respond to the rigidity of the polymeric templates on which they proliferate and differentiate by increasing their deposition of mineral content with increasing stiffness of the templates (Hung et. al., 2013).

The majority of bone defects occur at load bearing sites, therefore it is imperative for graft substitutes to maintain their structural and mechanical integrity from surgical implantation until bone fusion and healing is complete. As this graft substitute design is fabricated from biocompatible and biodegradable poly-lactic acid (PLA) polymer which as a degradation lifespan of approximately 1-2 years, the degradation profile may be controllable to match the rate of tissue in-growth based on the patient's specific needs (Liu and Ma, 2004; Mikos, 2006; Puppi et al., 2010; Pan and Ding, 2012). The graft substitute will degrade slowly as bone cells proliferate and mineralize to heal while still fostering a mechanically sound environment to sustain physiological weight bearing loads. This demonstrates the versatility of employing bioresorbable polymers as materials for the fabrication of bone graft substitutes (Ikada 2006; Polo-Corrales et. al., 2014). Therefore, future work would encompass characterizing the degradation profile of the graft substitute design to evaluate the osseointegration of the implanted tissue construct and to monitor the fusion of the bone end gaps of critically sized defects using appropriate large animal preclinical models (Reichart et. al., 2009). Applied clinically, this graft substitute design, which can be mechanobiologically optimized has the potential to overcome the current gold standard of autograft and allograft-associated limitations regarding critically sized bone grafting procedures.

## 5. CONCLUSIONS

To our knowledge, there are no synthetic graft substitutes that meet the mechanical and biological criteria that need to be satisfied to achieve long-term healing of critically sized segmental bone defects. We have developed a 3D printed bioresorbable graft substitute which serves as a viable candidate for the fusion of segmental bone end gaps, while also maintaining load bearing physiological functions to facilitate bone healing. The graft substitute comprises of a spatially graded two-layer skin and core structure to mimic native bone tissue where its local variations of porosity, pore size, and moduli at the skin and core layers can be readily customized to accommodate the size and location of segmental bone to be replaced, thereby producing patient-specific graft substitutes to orchestrate the regeneration of bone for the healing of load bearing segmental bone defects.